\documentclass[aps,prd,twocolumn,groupedaddress]{revtex4-2}

\newcommand\bmth[1]{\mbox{\boldmath $#1$}}
\newcommand\eqref[1]{(\ref{#1})}
\newcommand{\df}{{\mbox{\rm d}}}

\newcommand{\smfrac}[2]{{\textstyle{#1\over#2}}}
\def\half{\smfrac{1}{2}}

\begin{document}

\title{Comment: Exact vacuum solution with Hopf structure in general relativity}

\author{M. A. H. MacCallum}
\email[]{m.a.h.maccallum@qmul.ac.uk}
\affiliation{Queen Mary University of London}

\date{\today}

\begin{abstract}
  Harada's recently announced Kerr-Schild solution of Einstein's
  equations in general relativity [Phys. Rev. D {\bf 112}, 024020
    (2025)] is identified as isometric with one of the well-known NUT
    solutions.
\end{abstract}

\maketitle

 In a recent paper \cite{Har25}, Harada derived a Kerr-Schild vacuum solution of
 Einstein's equations, based on a  Hopf fibering, in the form 
 \begin{eqnarray} \label{Harada}
   &&\df s^2 = -2\df u\, \df v +\df x^2+\df y^2 \\
 \nonumber
   &&+H
       [\df v  - \frac{ux+by}{u^2+b^2} \df x
     -\frac{uy-bx}{u^2+b^2}\df y +\frac{x^2+y^2}{2(u^2+b^2)} \df u]^2.
 \end{eqnarray}
 where $H=Nu/(u^2+b^2)$.
Harada showed that this metric is of Petrov type D.  All Petrov type D
vacuum solutions are known \cite{Kin69}. The solutions formed one of
the first applications \cite{KarAma81} of the Cartan-Karlhede
procedure for characterization of spacetimes described in
\cite{SteKraMac03}, chapter 9. There is a useful table summarizing the
results in \cite{Ama84}. The files for the CLASSI software (see
\cite{MacSke94,Mac18}) used in
creating that table are freely available with the software.  Here I
use those files to identify which of the known solutions \eqref{Harada} is.

To apply the classification methods, the first step is choice of a suitable canonical tetrad.
 Debever \cite{Deb74} showed that Kerr-Schild spacetimes with a geodesic,
 diverging, and shearfree null vector $k_{\mu}$ can be written (using the
 conventions of \cite{SteKraMac03} and interchanging the names $u$ and $v$)  as
\begin{equation}
\label{eq:28.34}{\df}s^2=2({\df}\zeta \,{\df}\bar \zeta -{\df}u\,%
{\df}v)+2S({\df}v + \overline{Y}{\df}\zeta +Y{\df}\bar \zeta +Y\overline{Y}%
{\df}u)^2
\end{equation}
with $\zeta = (x+iy)/\sqrt{2}$. This is of the same form as (\ref{Harada}) if
$a=u+ib$, $Y=-\zeta/a$ and $S=H/2$. Following  \cite{SteKraMac03}, a suitable
tetrad for \eqref{eq:28.34} is
\begin{eqnarray}\label{28.33}
\bmth{\omega}^1& =&  \df \zeta +Y \df u, \quad \bmth{\omega}^2=
\df\bar{\zeta}+\overline{Y} \df u,  \label{eq:28.33} \\
\bmth{\omega}^3& =& \df v+ \overline{Y} \df \zeta+Y \df\bar{\zeta}+Y \overline{Y}
 \df u, \quad \bmth{\omega}^4=\df u- S\bmth{\omega}^3.\nonumber
\end{eqnarray}

Using \eqref{eq:28.33}, CLASSI readily confirmed that \eqref{Harada}
is a Petrov type D vacuum solution.  In the coordinates of
\eqref{Harada}, Harada noted the Killing vectors $\partial_v$ and
$y\partial_x-x\partial y$, but CLASSI reports the metric has 4 Killing
vectors. This could have easily been foreseen since the invariants
contain only one essential coordinate.

Harada's solution has (in Newman-Penrose \cite{NewPen62} notation)
$\Psi_2=-N/2(u+ib)^3$ as the only nonzero component of the Riemann
tensor, and an invariant from the derivative of the Weyl tensor
\begin{equation}
 C^{abcd;e} C_{abcd;e}       = 45N^3u/2(u^2+b^2)(u+ib)^8.
\end{equation}
Using {\AA}man's files and coordinates, Kinnersley type I metrics with $C=0$ have the
invariants $\Psi_2=-m/(r+i\ell)^3$ and $C^{abcd;e} C_{abcd;e} =
180m^3r/(r^2+\ell^2)(r+i\ell)^8$. Clearly these agree with the Harada
solution if we identify $r=u,  b=\ell, N=2m$.

The other cases in {\AA}man's table which have 4 Killing vectors have
different invariants, not matching those of Harada's
solution. Specifically, Kinnersley Type IIA with $a=0$ has
$\Psi_2=-(m+i\ell)/(r+i\ell)^3$, 
 Kinnersley Type IV A with nonzero $a$ has extra terms in $C^{abcd;e}
 C_{abcd;e}$ and Kinnersley Type IV B has a real $\Psi_2$.
 There is a error in the table for Type II.BCD with $b=0$
 which has only 2, not 4, Killing vectors.

Hence from these comparisons one finds that Harada's solution is
locally isometric to one of the well-known Taub-NUT solutions, the one
with a zero two-space curvature parameter. (This parameter is denoted
$\mu_0$ in \cite{NewTamUnt63} and $\epsilon$ in \cite{GriPod09}.)

Starting from the Debever form \eqref{eq:28.34} of the metric enables
a concise way to present coordinate transformations from
\eqref{Harada} to standard forms of the NUT metric. In
\eqref{eq:28.34} substitute $\zeta=az, \df\zeta=a \df z +z \df u$. The
metric then becomes
\begin{equation}
 \df s^2 =  -2 \df u \, \sigma +2(u^2+b^2)\df z\df{\bar{z}} +H\sigma^2
\end{equation}
where
\begin{equation}
  \sigma=
(\df v - a\bar{z}\df z
-\bar{a}z\df{\bar{z}}-z\bar{z}\df u)
\end{equation}
Substituting $a=u+ib$ and $z=(P+iQ)/\sqrt{2}$ (so that $P=(ux+by)/(u^2+b^2),
Q=(uy-bx)/(u^2+b^2)$) into $\sigma$ one finds
\begin{eqnarray}
  \sigma &=& \df v -u(P\df P+Q\df Q)-\half(P^2+Q^2)\df u\\
\nonumber  &&+bP\df Q-bQ\df P \\
\nonumber & =& \df v -\half \df((P^2+Q^2)u)+bP\df Q-bQ\df P.
\end{eqnarray}
Making the variable transformation $v=w+\half ((P^2+Q^2)u)$ the
metric becomes
\begin{equation}\label{NUT1}
 \df s^2= -2 \df u\ \sigma +(u^2+b^2)(\df P^2+\df Q^2) +H\sigma^2,
\end{equation}
with $\sigma = \df w +b(P\df Q-Q\df P)$. This is the original NUT form for
the case $\mu_0=0$, with renamed parameters
and coordinates. \cite{NewTamUnt63} gives the Killing vectors Harada
did not find as $\partial _P-bQ\partial_w$ and  $\partial_Q +
bP\partial _w$.

If one substitutes $w=V+\int \df u/H$ and
introduces polar coordinates $(\rho,\,\phi)$ in the $(P,\,Q)$ plane the metric becomes
\begin{equation}\label{NUT2}
  -\df u^2/H+(u^2+b^2)(\df \rho^2+\rho^2\df\phi^2) +H(\df V+b\rho^2 \df\phi)^2,
\end{equation}
which (with $V=t$ and $H=-f$ there) is the form used in
\cite{GriPod09}, where \S 12.3.2 analyses properties such as the
conformal structure. In the coordinates of \eqref{NUT2} the Killing
vectors that Harada did not find are
$\cos \phi\, \partial_\rho  - \sin \phi(1/\rho \,\partial_\phi+b\rho \partial_V)$ and
$\sin \phi \,\partial_\rho   + \cos \phi(1/\rho \,\partial_\phi+b\rho \partial_V)$.

I am grateful to their authors for informing me that the same
identification had been given in \cite{AyoFloHas25} and \cite{Mao25}.
\bibliography{harada}
\end{document}